\documentclass[11pt,twoside]{article}

\usepackage{asp2014}

\aspSuppressVolSlug
\resetcounters

\begin{document}

\title{Revealing Predictive Maintenance Strategies from Comprehensive Data Analysis of ASTRI-Horn Historical Monitoring Data}

\author{Federico~Incardona,$^1$ Alessandro~Costa,$^1$ Giuseppe~Leto,$^1$ Kevin~Munari,$^1$ Giovanni~Pareschi,$^2$ Salvatore~Scuderi,$^3$ and Gino~Tosti$^4$ for the ASTRI project$^*$}
\affil{$^1$INAF, Osservatorio Astrofisico di Catania, I-95123 Catania, Italy; \email{federico.incardona@inaf.it}}
\affil{$^2$INAF, Osservatorio Astronomico di Brera, I-23807 Merate, Italy}
\affil{$^3$INAF, Istituto di Astrofisica Spaziale e Fisica Cosmica, I-20133 Milano, Italy}
\affil{$^4$Universi\`a di Perugia, Istituto di Astrofisica Spaziale e Fisica Cosmica, I-20133 Milano, Italy}
\affil{$^*$\url{http://www.astri.inaf.it/en/library/}}

\paperauthor{Federico~Incardona}{federico.incardona@inaf.it}{0000-0002-2568-0917}{INAF}{Osservatorio Astrofisico di Catania}{Catania}{}{I-95123}{Italy}
\paperauthor{Alessandro~Costa}{alessandro.costa@inaf.it}{}{INAF}{Osservatorio Astrofisico di Catania}{Catania}{}{I-95123}{Italy}
\paperauthor{Giuseppe~Leto}{giuseppe.leto@inaf.it}{}{INAF}{Osservatorio Astrofisico di Catania}{Catania}{}{I-95123}{Italy}
\paperauthor{Kevin~Munari}{kevin.munari@inaf.it}{}{INAF}{Osservatorio Astrofisico di Catania}{Catania}{}{I-95123}{Italy}
\paperauthor{Giovanni~Pareschi}{giovanni.pareschi@inaf.it}{}{INAF}{Osservatorio Astronomico di Brera}{Merate}{}{I-23807}{Italy}
\paperauthor{Salvatore~Scuderi}{salvatore.scuderi@inaf.it}{}{INAF}{Istituto di Astrofisica Spaziale e Fisica Cosmica}{Milano}{}{I-20133}{Italy}
\paperauthor{Gino~Tosti}{gino.tosti@unipg.it}{}{Universit\`a di Perugia}{Dipartimento di Fisica e Geologia}{Perugia}{}{I-06123}{Italy}



\begin{abstract}
Modern telescope facilities generate data from various sources, including sensors, weather stations, LiDARs, and FRAMs. Sophisticated software architectures using the Internet of Things (IoT) and big data technologies are required to manage this data. This study explores the potential of sensor data for innovative maintenance techniques, such as predictive maintenance (PdM), to prevent downtime that can affect research. We analyzed historical data from the ASTRI-Horn Cherenkov telescope, spanning seven years, examining data patterns and variable correlations. The findings offer insights for triggering predictive maintenance model development in telescope facilities.
\end{abstract}



\section{Introduction}
ASTRI-Horn \citep{2016SPIE.9906E..5TP} is an Italian telescope designed to detect gamma rays as a prototype of the Small-Sized Telescopes of the Cherenkov Telescope Array (CTA) project \citep{CTA2018}. It was developed in the context of ASTRI \citep[``Astrofisica con Specchi a Tecnologia Replicante Italiana'',][]{ASTRI:2013heb}, a project that aims at deploying an array of nine ASTRI-like telescopes (named the ASTRI Mini-Array) at the Observatorio del Teide in Tenerife (Spain) for the study of high-energy gamma-ray emissions from astrophysical sources \citep{2022JHEAp..35....1V, 2022JHEAp..35...52S}. As a whole, the system will be an Internet of Things environment, requiring a complex software architecture to address data collection and storage \citep{incardona2022monitoring, Costa:2021c8}.

\section{Telescope activity windows}
The ASTRI-Horn prototype was installed in 2014 at the observing station of the INAF Astrophysical Observatory of Catania (Serra La Nave, Mt. Etna).


We analyzed the monitoring time series collected by the Telescope Control Unit (TCU) and the Telescope Health Control Unit (THCU) \citep{2016SPIE.9913E..3LT}, which are in charge of monitoring the ASTRI-Horn housekeeping data, and by a Weather Station (WS) located in the proximity of the telescope.

The data spans from June 2017 to October 2023, but we selected only intervals when the motors were active. We obtained 1307 intervals, for which we estimated an average time duration of about 90 minutes. Fig.\ref{ex_fig1} reports the interval amplitude distribution in seconds.
\articlefigure[width=0.55\textwidth]{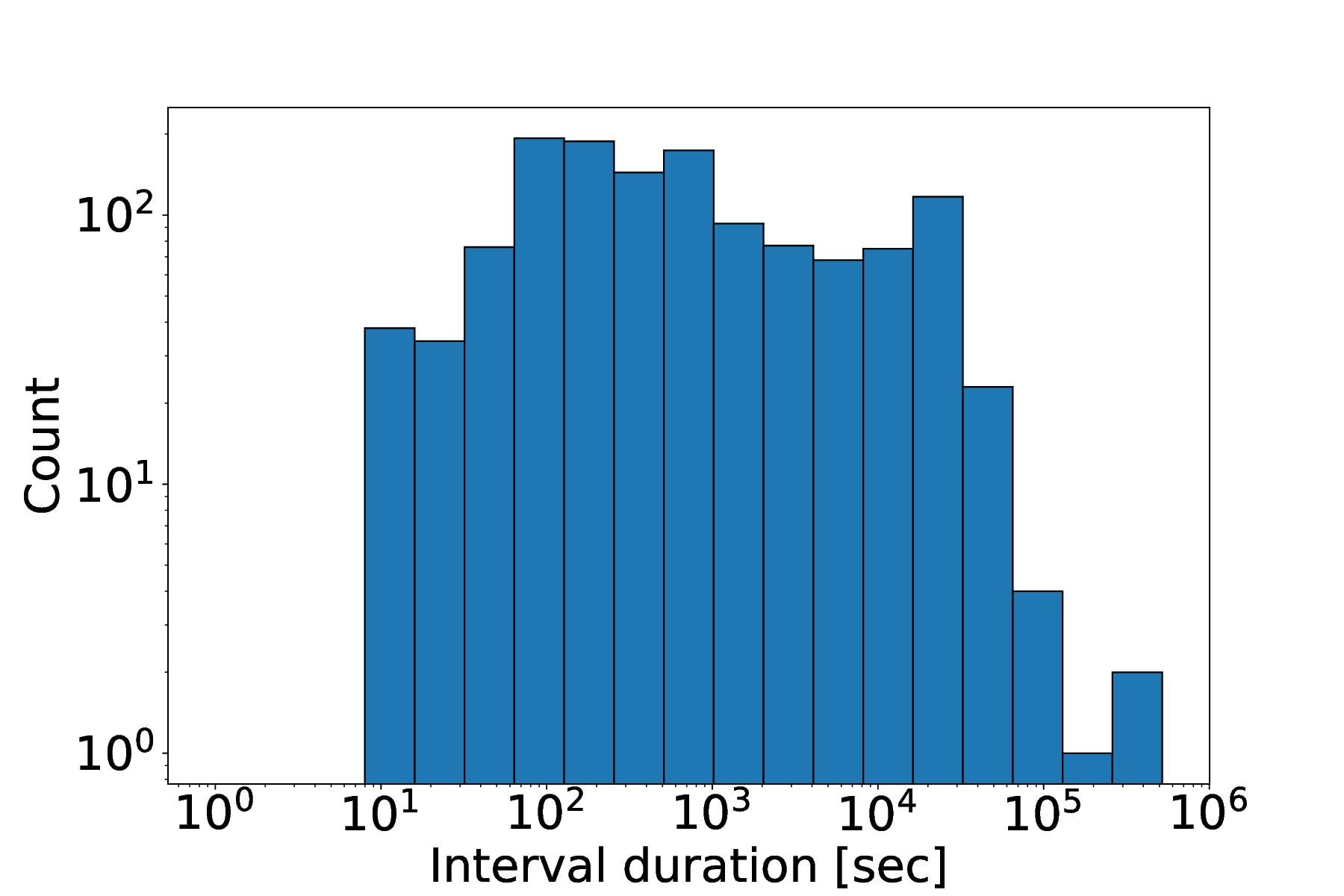}{ex_fig1}{Distribution of the activity windows duration of the ASTRI-Horn telescope since 2017.}

\section{Variables correlation and principal component analysis}
The set of available monitored properties for TCU, THCU, and WS was composed of more than 200 time series, each of which was sampled with a frequency of about 15 seconds, even if, in many cases, the sampling rate was not uniform.

We focused on a subset of monitored properties composed of 19, 27, and 4 variables, respectively, for TCU, THCU, and WS. These features report information about the currents, voltages, phases, positions, temperatures, torques, and statuses of the telescope components, as well as the environmental temperature, humidity, wind speed, and solar radiation.
\articlefigure[width=0.7\textwidth]{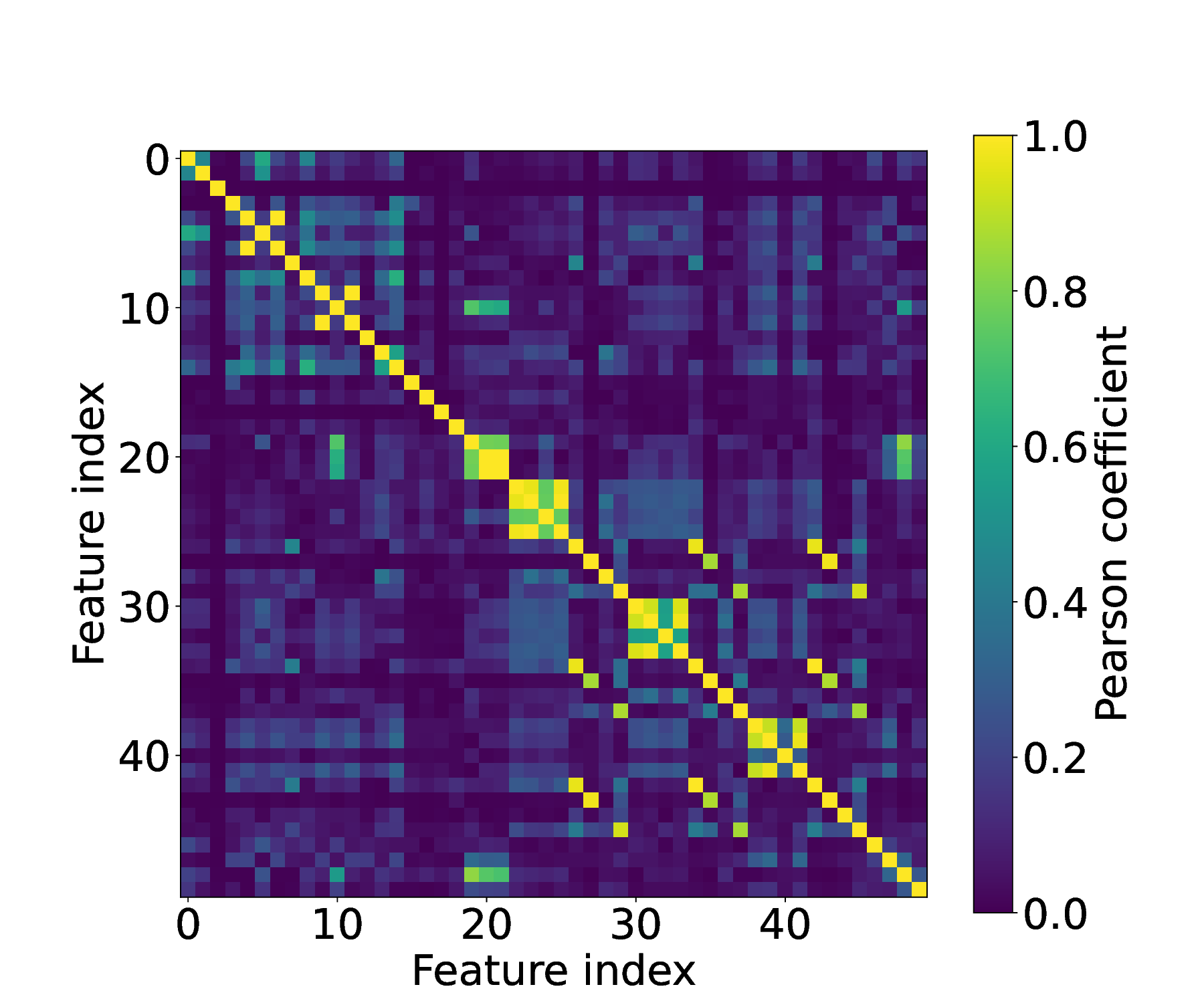}{ex_fig2}{Correlation matrix for a subset of TCU, THCU, and WS variables collected from 2017 to 2023 and resampled at one minute.}

To reduce the number of data points, we resampled each active interval of the dataset at a rate of 1 minute and computed the correlation matrix (Fig.\ref{ex_fig2}). We also used larger time scales (1, 6, 12 hours, one day, and one week) to resample the dataset and every time we computed the correlation matrix.

We found tens of correlations with the larger values of the Pearson coefficient\footnote{The Pearson coefficient $r$ is the ratio between the covariance of two variables and the product of their standard deviations and measures the linear correlation between two variables \citep{10.5555/59551}.} ($r > 0.9$) at every time-scale. On the time scale of the week, we observed a moderate correlation with $r \approx 0.53$ between the Azimuth master motor current and encoder initialization status. This correlation strengthened to $r \approx 0.75$ when we focused on the time frame just before the encoder replacement, which happened in April 2022 due to wear. This finding suggests a potential wear signature in the time series data that warrants further investigation.

Finally, we performed a Principal Component Analysis \citep[PCA,][]{10.5555/59551} on the dataset. The \textit{scree plot}\footnote{A scree plot is a graph of eigenvalues against the corresponding principal component number \citep{10.5555/59551}.} reported in Fig.\ref{ex_fig3} shows that about 16 principal components retain about 80\% of the total dataset variance.
\articlefigure[width=0.77\textwidth]{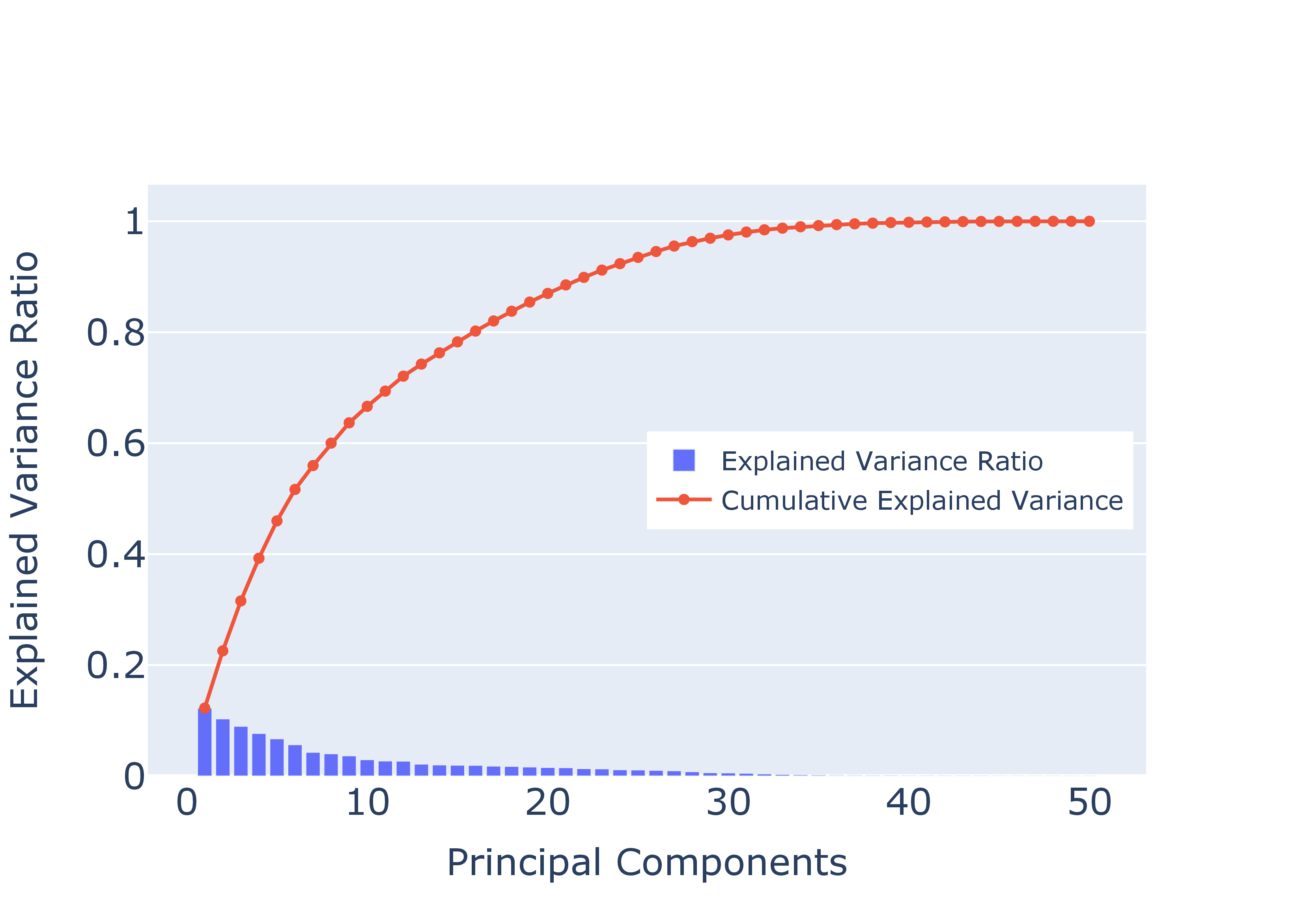}{ex_fig3}{Amount of dataset variance explained by each principal component (bars) and cumulatively (line with markers).}

\section{Insights for predictive maintenance models}
Predictive maintenance \citep[PdM, ][]{MOBLEY200299} is a data-driven approach to maintenance that improves the life cycle of a system by predicting \textit{just in time} potential equipment malfunction, reducing its downtime. Such an innovative approach can be applied to telescope facilities \citep{incardona2022failure} in combination with some of the most advanced Machine Learning models \citep{10.1007/978-3-031-34167-0_41}.

We aim to develop a model that constantly compares, for each target feature, the actual measured values and the predicted ones and raises an alarm every time a deviation from the model is detected. We will exploit the stronger correlations and principal components to reduce model overfitting. Confirmed correlations due to wear will be used to identify the correct prediction window for that kind of damage. Finally, we plan to repeat this analysis by considering a wider set of monitored properties.

\acknowledgements This work was conducted in the context of the ASTRI Project. We gratefully acknowledge support from the people, agencies, and organizations listed here: \url{http://www.astri.inaf.it/en/library/}. This paper went through the internal ASTRI review process. This work is also supported by the ``Istituto Nazionale di Astrofisica'' (INAF) Mini-Grant.

\bibliography{P402}  


\end{document}